# ALL-OPTICAL NONLINEAR ACTIVATION FUNCTION FOR PHOTONIC NEURAL NETWORKS

MARIO MISCUGLIO[1], ARMIN MEHRABIAN[1], ZIBO HU[1], SHAIMAA I. AZZAM[2], JONATHAN GEORGE[1], ALEXANDER V. KILDISHEV[2], MATTHEW PELTON[3], VOLKER J. SORGER[1*]

[1] *Department of Electrical and Computer Engineering, George Washington University, Washington, DC 20052, USA*
[2] *School of Electrical & Computer Engineering and Birck Nanotechnology Center, Purdue University, West Lafayette, Indiana, 47907, USA*
[3] *Department of Physics, UMBC (University of Maryland, Baltimore County), Baltimore, Maryland 21250, USA*

**With the recent successes of neural networks (NN) to perform machine-learning tasks, photonic-based NN designs may enable high throughput and low power neuromorphic compute paradigms since they bypass the parasitic charging of capacitive wires. Thus, engineering data-information processors capable of executing NN algorithms with high efficiency is of major importance for applications ranging from pattern recognition to classification. Our hypothesis is therefore, that if the time-limiting electro-optic conversion of current photonic NN designs could be postponed until the very end of the network, then the execution time of the photonic algorithm is simple the delay of the time-of-flight of photons through the NN, which is on the order of picoseconds for integrated photonics. Exploring such all-optical NN, in this work we discuss two independent approaches of implementing the optical perceptron's nonlinear activation function based on nanophotonic structures exhibiting i) induced transparency and ii) reverse saturated absorption. Our results show that the all-optical nonlinearity provides about 3 and 7 dB extinction ratio for the two systems considered, respectively, and classification accuracies of an exemplary MNIST task of 97% and near 100% are found, which rivals that of software based trained NNs, yet with ignored noise in the network. Together with a developed concept for an all-optical perceptron, these findings point to the possibility of realizing pure photonic NNs with potentially unmatched throughput and even energy consumption for next generation information processing hardware.**

## 1. Introduction

The past decades have been marked by an exponential increase in demand for high speed and energy efficient computer architectures. The established microelectronics technology faces its biggest limitations with respect to handling real-time processing, large data volumes and complex systems. The exponential growth in transistor count and integration, which marked the previous century, cannot be pushed further, highlighting the twilight of Moore's law for CMOS technology. In this changing landscape, multiple companies and research groups are pursuing novel solutions to provide high-performance computing to the next level, using better performing algorithms and novel technological approaches, where the work presented here falls under the latter. A widely investigated solution is to replace general-purpose processors (von Neumann architecture) with more specialized and task-specific processors. Graphic Process Units (GPUs) or Field Programmable Gate Array (FPGAs), architectures, which are optimized for data computation and parallelism provide efficiencies for their specialized tasks up to an order of magnitude higher than generic-task CPUs [1-3]. However, these approaches still rely on electronic transport and are bound by the speed and power limits of the interconnects inside the circuits due to RC parasitic effects.

A subclass of algorithms which receives benefits when implemented in non von Neumann architecture is represented by Neural Networks (NN). NNs are designed to simultaneously process large arrays of data in order to learn representations of them with multiple levels of abstraction. These architectures consist of neurons that perform the following basic functions: 1) Interpret multiple incoming signals in a form of multiple arrays (e.g. images) through weighted addition (Multiply and Accumulate, MAC); 2) Apply a nonlinear (NL) activation function for discriminating the data; 3) Transmit the result to multiple destination neurons (i.e., fan-out).

Recently emerging Photonic Neural Networks (PNN) demonstrated the potential to increase computing speed by 2-3 orders of magnitude [4]. In order to implement the functions of the NN into a PNN, two classes of devices and their respective functions need to be engineered, the weighted sum and the NL activation. The weighted sum, as previously investigated [5], can be obtained by a combination of balanced photodetection, Mach-Zehnder interferometer, or ring resonator, which constitutes the weights of the signal, and y-junctions or MMI (Multimode interference coupler) for summation. The weighted addition in analog photonics, which is the equivalent of the MAC, uses optical interference, while the coherent electromagnetic waves propagate through the photonic integrated circuit (PIC), and offers MAC energy consumption that does not trade-off with MAC speed [6]. Moreover, photonic interconnects can have large bandwidth and low power consumption.

On the other hand, the absence of a straightforward and efficient NL in optics has severely limited its usefulness in DeepLearning computing. Several attempts [7-9] showed that a NL modulation of the optical signal could be achieved through Electro-optic Absorption Modulators (EOMs). These rely on the on-chip implementation of an electronic platform that is able to vary the optical effective mode index by tuning the voltage. Despite their relatively straightforward implementation and controllability, EOM results in substantial trade-offs in terms of speed and power efficiency of the otherwise intrinsically instantaneous transmission of the signal through the PIC.

Other approaches [10] consist of devices that would need first to convert an optical signal into an electrical signal by means of a detection mechanism and afterwards convert it back into an optical signal, a solution that would hamper the speed and cascadability of the network due to both the movement of charge carriers and noise (e.g. shot and thermal noise of a photodetector). For this reason, the implementation of fast and energy efficient all-optical nonlinearity becomes a key task for boosting the throughput of a neural network and consequently lowering the latency and power dissipation.

In this work, we aim to overcome these limitations by implementing novel devices and approaches based on light matter interaction (LMI) in assembly of nanoparticles in PICs. A first device prototype relies on a reversible transparency induced by a Fano resonance in a plasmon-exciton system. This system consists of a semiconductor nanocrystal, i.e. quantum dot (QD), within two gold nanorods. We also studied the NL response induced by reverse saturable absorption in films of buckyballs ($C_{60}$). The optical response of the two hybrid systems shows a strong non-linearity suitable for being employed as activation functions in PNN circuits as investigated here. We find the optimal configurations that optimized the mode coupling between the signal transmitted in a waveguide and the hybrid structures using numerical tools. Finally, a NN architecture that exploits the proposed activation functions has been emulated on an open source machine-learning framework, i.e. Tensorflow. We then compared the performances in terms of speed and accuracy of the proposed system with others, which employ conventional nonlinear functions (e.g. Tanh, Sigmoid, ReLu), for a specific DeepLearning tasks. Ultimately, this work points towards further use of NL activation functions of optically triggered devices, integrated with silicon photonics, for the next generation of photonic processors.

## 2. Discussion

As a first step in our study we investigate the LMIs in hybrid systems consisting of a pair of gold nanorods separated by a 10-nm gap that contains a single CdSe Quantum Dot (QD) Fig. 1a. Despite the apparent complexity, these quantum-assemblies could be either fabricated through a bottom-up approach, i.e. colloidal synthesis, as well as fabricated through a combination of top-down and bottom-up approach, by combining electron beam lithography and guided colloidal deposition [11]. As reported before [12-15], and illustrated in Fig. 1a, the hybrid system could be modeled as two coupled oscillators, one representing the dipole of the QD transition, and the other representing the dipole of the plasmon. The surface plasmon is driven by the incident field, and the QD is driven due to its coupling with the plasmon. Destructive interference between the two oscillators leads to cancelation of the optical response of the coupled system, known as an induced transparency or a Fano resonance.

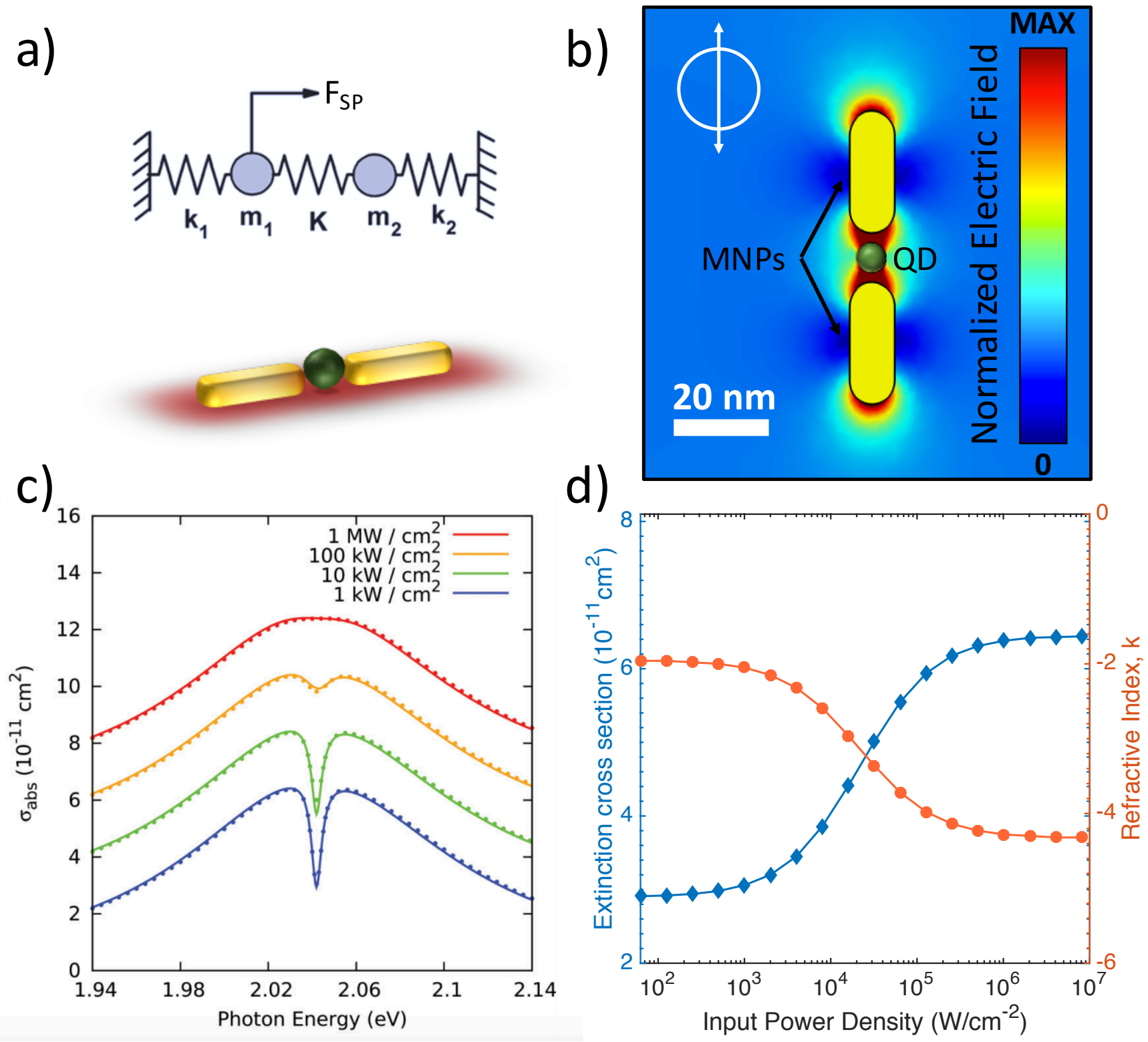

Fig.1. Physical modeling and material characterization: a) Phenomenological description system. Two coupled oscillators that provides a classical analogue of the plasmon-exciton coupling induced transparency in the represented system and schematic representation of the MNPs/QD system consisting of a single quantum dot (QD) between a pair of gold nanoparticles (MNP). b) Normalized electric field distribution of the hybrid system computed for a 2.04 eV impinging plane wave. Scale bar 20 nm. c) Calculated absorption spectra for a system illustrated in the inset of (a) taken from [16]. The different curves show the response of the system for different amounts of energy in an incident laser pulse. d) Extinction cross section (left y-axis) and imaginary part of the refractive index (right y-axis) of the MNPs/QD system, as function of the input power density.

Figure 1b displays the normalized electric near field distribution computed at 2.04 eV, in resonance with the gold rods of the proposed MNPs/QD system. The model parameters for the metal MNPs/QD system are obtained from FDTD simulations [14]. The complex refractive index of Gold and CdSe QD were obtained from refs [17] and [18]. The large electric field in the feed-gap of a gold optical antenna [19], generated by the plasmonic dipole resonances of the nano-rods, completely engulfs the CdSe QD, inducing the coupling between the plasmon and the QD that ultimately gives rise to the Fano resonance, which shows the obtained absorption spectra of the hybrid system for different incident optical powers, taken from ref [14] (**Fig. 1c**). The Fano resonance, or transparency dip, is apparent for low incident optical power; this linear response has recently been confirmed by experimental results [12].

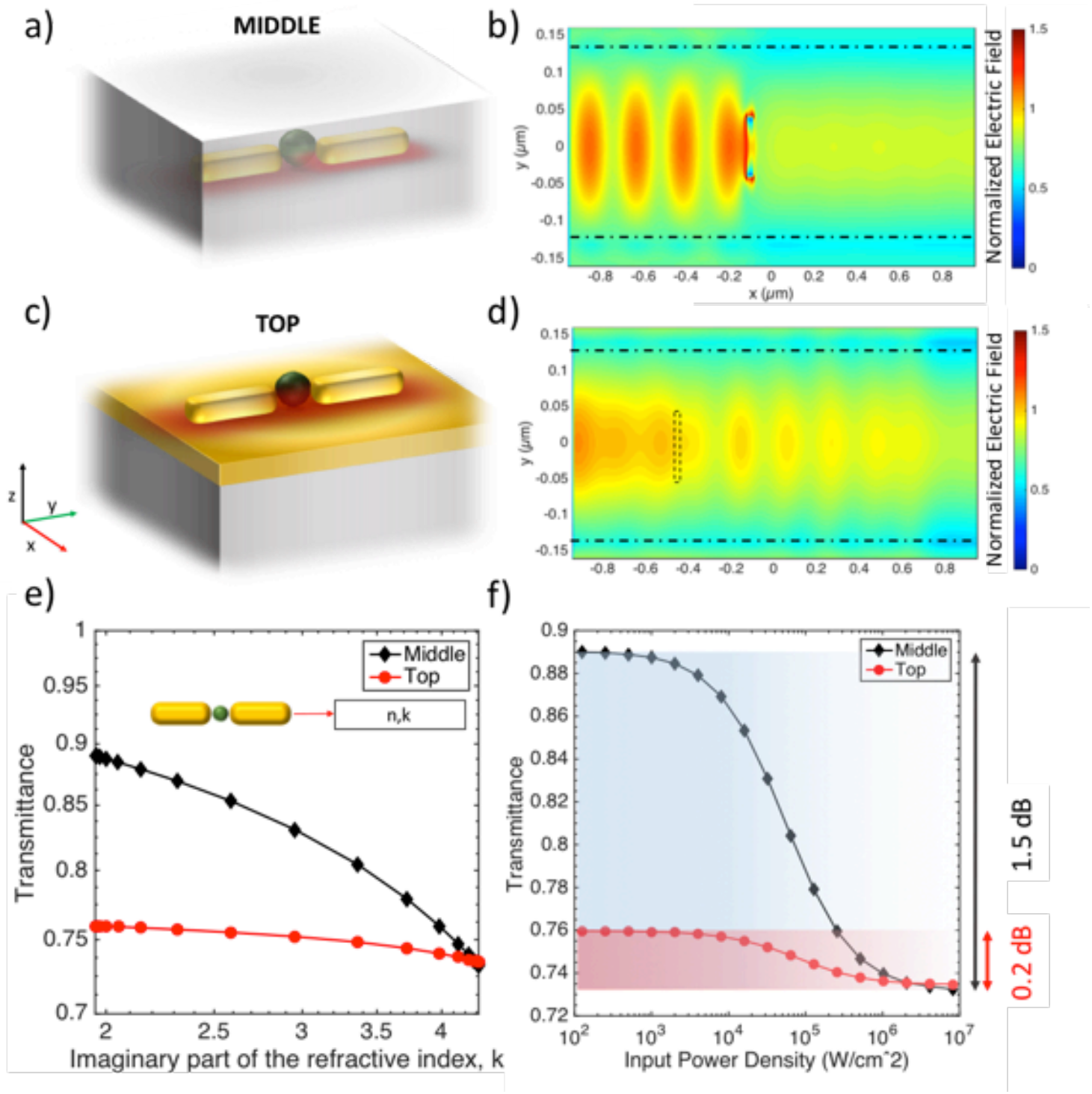

Fig. 2. Hybrid-Nanoparticle (NP) waveguide integration for an all-optical nonlinearity for photonic neural networks: a-c) Schematic representation of waveguide and MNPs/QD system coupling. The MNPs/QD system placed in the middle (a) and top (c). b-d) Normalized electric field distribution for a middle horizontal cut plane of the waveguide considering the assembly being placed in the middle (b) and on top (d), for its maximum absorption (highest *k*). c) Computed transmittance of the waveguide as function of the tunable absorption of a MNPs/QD system placed in the middle (Black solid line) and on top of the waveguide (Red solid line). d) Computed waveguide transmittance as function of the input power density and respective nonlinear modulation ranges in dB.

As the incident power increases, the Fano resonance dip disappears, due to saturation of the QD transition. Thus, the amount of energy dissipation in the coupled system is a NL function of the input power. This nonlinearity can be seen in the extinction cross section (left y-axis) spectra of the assembly computed at 2.04eV, as function of the input power density (Fig. 1d), suggesting a narrow spectral operation band. In order to later embed the system in a photonic waveguide, the imaginary part of the refractive index of the assembly as function of the input power density, was derived using a simplified geometry and Mie Scattering model [20] (right y-axis), which manifests a specular behavior.

We proceed further in our study by embedding the MNPs/QD system in a silicon photonic waveguide and studying their interaction through a finite-difference time-domain (FDTD) solver. The previously described physical model was investigated in two different device configurations, placing the assembly in the middle and on top of the waveguide, as schematized in **Fig. 2 a** and **c**, respectively. When placed on top, a 50 nm thin gold layer was used for enhancing the mode overlap between the assembly and the transmitted waveguide signal.

**Fig. 2b** and **d**, maps the normalized electric field travelling within the waveguide in the horizontal plane. Here, the absorbance of the assembly is set to its maximum; i.e., the imaginary part of the refractive index, *k*, is maximized. In this case, the overall transmitted power is similar for both configurations. As shown in **Fig. 2e-f**, we evaluate these configurations by analyzing the modulation range of the transmitted power as a function of the imaginary part of the refractive index (**e**), which, for its part, depends on the input power transmitted in the waveguide (**f**). It may be observed that in both the configurations, the sole presence of an assembly produces a tangible NL modulation as function of the input optical power, while showing a sigmoidal shape. As expected, the configuration with the MNPs/QD system in the middle gives rise to a larger modulation range, reaching ~1.5 dB compared to the 0.2 dB obtained with a particle placed on top.

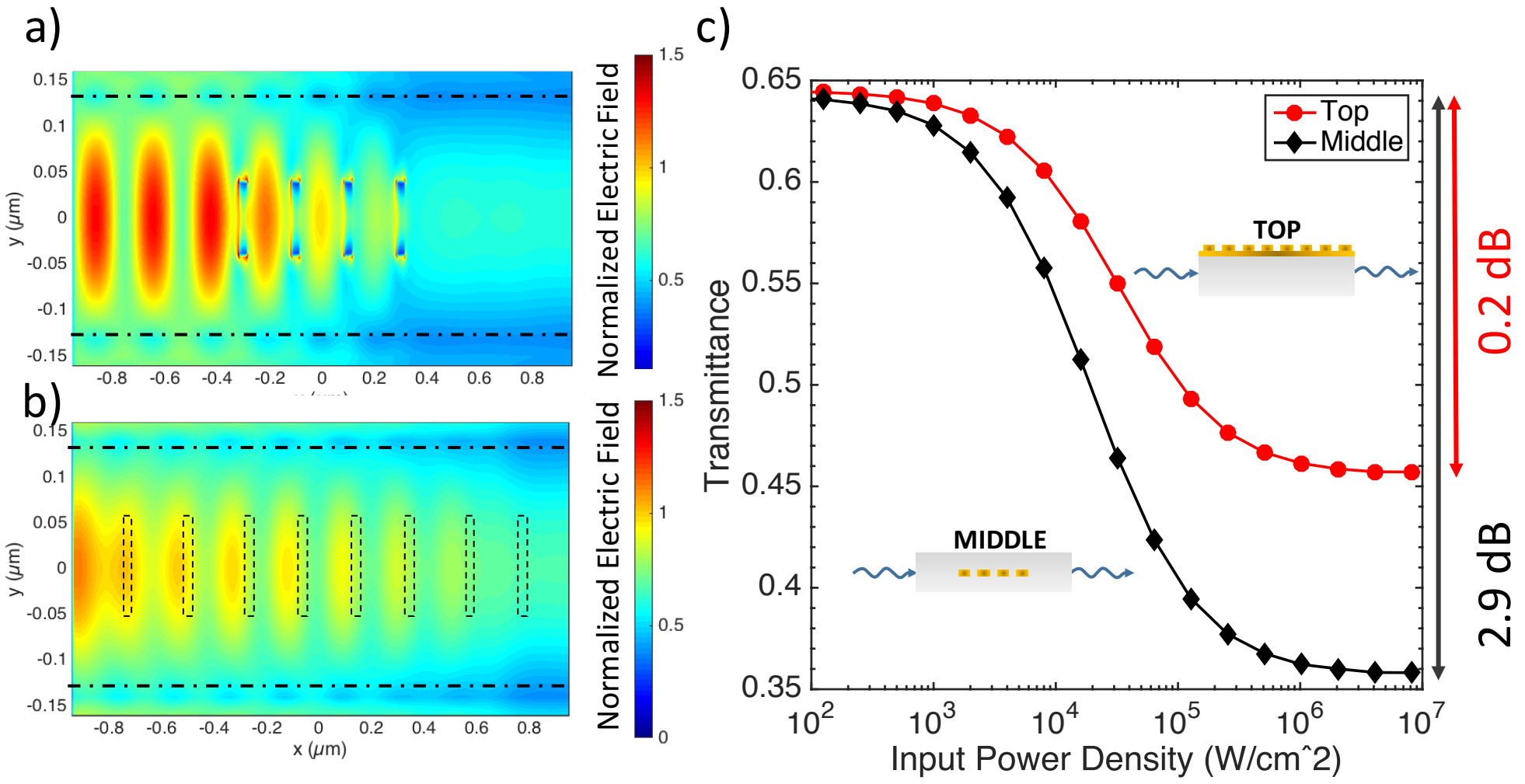

Fig. 3. Engineering the all-optical nonlinearity via array: a-b) Normalized electric field distribution for a middle horizontal cut plane of the waveguide, considering an array of closely spaced hybrid-structures, being in the middle (a) and on top (b) of the waveguide, for their maximum absorption (largest k). (c) Transmittance as function of the input power density.

Proceeding further with our analysis, in order to enhance the modulation range and consequently to introduce larger NLs in the transmitted signal, we consider the system composed by arrays of MNPs/QD assemblies that are distributed either on top or in the middle of the waveguide. Similarly to what is shown for single assemblies, Fig. 3(a,b) shows the electric field distribution in the horizontal cut-plane of the waveguide, when either 4 or 8 assemblies are placed in the middle and on top. The assemblies are spaced with a 200 nm pitch in both configurations. In both the configurations, the electric field transmitted at the end of the waveguide is significantly attenuated, when the assemblies retain their maximum *k*. More specifically, the modulation depth now approaches 50% for the top configuration and 65% for the middle configuration. Fig. 3(c) shows the NL modulation range of the proposed devices as function of the input power density. The NL can affect the transmitted power by more than 2 dB in the top array configuration and almost 3 dB in the middle array configuration.

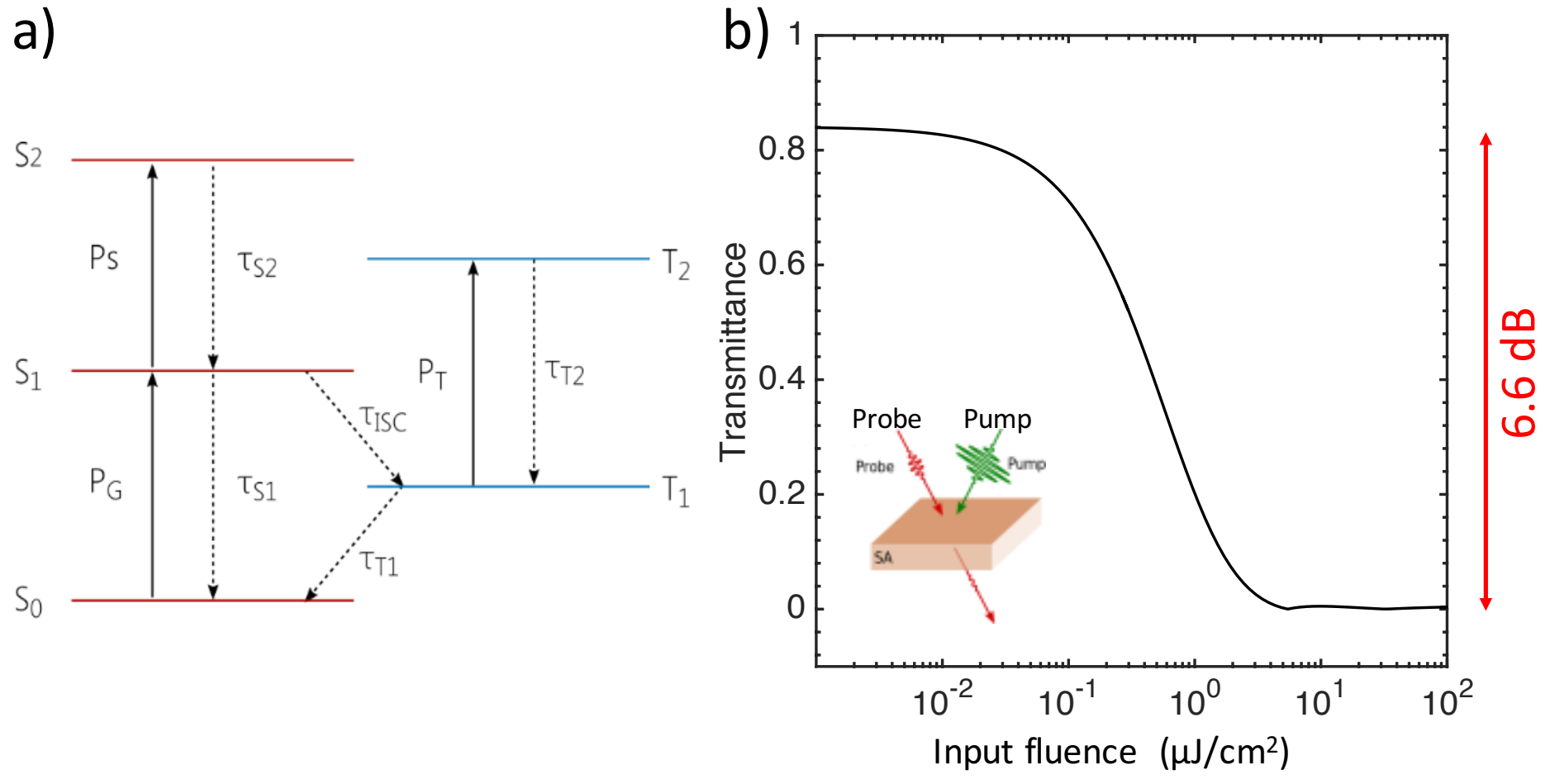

Figure 4. Optical NL in a Reverse Saturable Absorber. Nonlinear transmission of a reverse saturable absorber made of high-concentration $C_{60}$ in a PVA host thin film. (**a**) The simplified band-diagram of the reverse saturable absorber modeled by a five-level system. (**b**) Transmission vs. input fluence at 532 nm. Pump/Probe parameters: full-width at half maximum = 1 ps and of the probe = 5 fs. Both pump and probe are centered at a wavelength of 532 nm. The lifetimes, $\tau_{S1}$ = 30 ns, $\tau_{T1}$ = 280 μs, $\tau_{ISC}$ = 1.2 ns, $\tau_{S2} = \tau_{T2}$ = 1ps.

Another possibility for inducing optical NL in a PIC is exploiting reverse saturable absorption (RSA). RSA is a property of a material for which the absorption increases as the light intensity increases. The response of the coupled MNPs/QD assemblies is an ultra strong, "artificial" RSA that arises from the Fano interference between the components. However, a broad range of materials show "natural," albeit weaker, RSA, such as organic compounds [21,22], heavy metals [23], and clusters of metallic particles [24]. In analogy to the discussion before, we next consider the use of $C_{60}$ (Buckminsterfullerene or Buckyball) dispersed in polyvinyl alcohol (PVA) as a reverse saturable absorber to obtain an intensity-dependent transmission as a NL activation function. The band diagram of the majority of reverse saturable absorbing materials can be approximated by five energy levels representing two singlet and one triplet states, as depicted in Fig. 4a. We use rate equations to model the transitions among the different energy levels and to capture the carrier kinetics in the system and model the nonlinear light-matter interaction. Full details of the model can be found in Ref. [25]. The model is integrated with a finite-difference time-domain (FDTD) solver using an auxiliary differential equation approach to perform full-wave multiphysics analysis of the structures using the same structured grid and time-stepping.

A pump-probe analysis is used to extract the nonlinear transmission where a strong pump signal first illuminates the structure followed by a weak probe signal to probe the system response (Fig. 4a). The respective lifetimes are taken from optical characterization [22]. The concentration of the $C_{60}$ molecules is chosen to be 10 mM to provide sufficient absorption from a thin film. The transmission of 1 μm-thin film of $C_{60}$ in PVA (refractive index ~ 1.45) is calculated as a function of the input fluence and shown in Fig. 4b. The linear transmission of the film is 0.84, and the saturation fluence is about 0.67 μJ/cm$^2$. We can also conclude that the NL modulation associated with the input power is approximately 7 dB.

After obtaining a small library consisting of 3 different NL optical responses, we introduce these as nonlinear activation functions on a standardized NN training set, MNIST classifiers of handwritten digits.

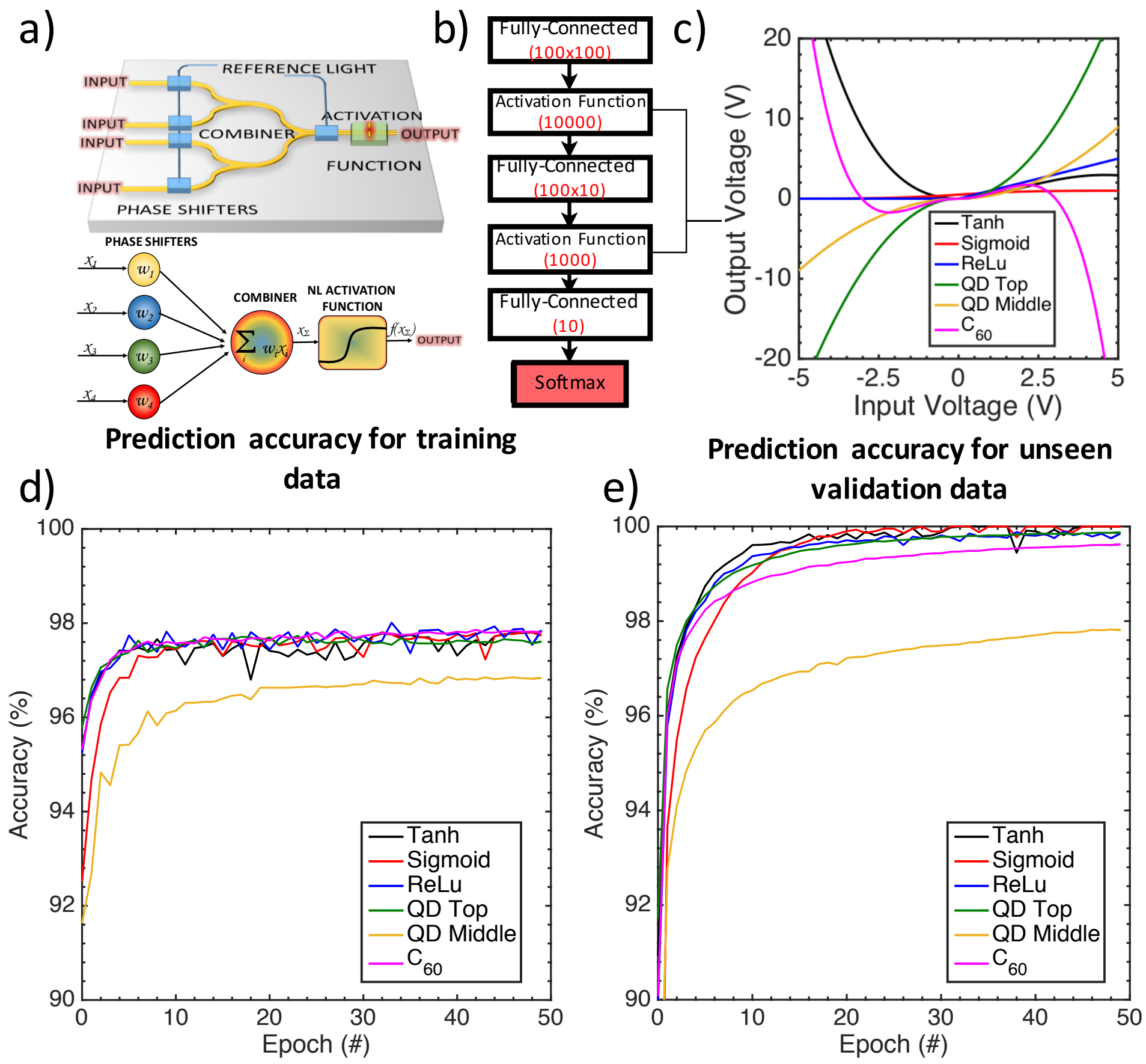

Fig. 5. Design of an AO neuron and evaluation of the Neural Network performance a) Schematic of an all-optical neuron. b) Representation of the emulated NN (b). c) Different activation functions, $V_{IN}$ to $V_{OUT}$, ReLU (light blue dashed line), Sigmoid (blue dashed line), hyperbolic tangent (dark blue line) and the 2 proposed NL function, top (solid orange) and middle (solid red) d-e) Prediction accuracy as function of epoch for the training (d) and validation (e) phase of the network.

Figure 5a shows the overall hardware implementation of the AONN using photonic integrated circuits. As previously reported [5,26], the input optical signals are weighted by phase shifters and integrated by photonic combiners. Hereby, the NL activation functions (AF) can be achieved either by means of induced transparency through plasmon-exciton coupling or by reverse saturable absorption of $C_{60}$. The proposed all-optical NL modulation of the signal presents several advantages over an electro-optical counter part such as no charge fluctuation-induced noise or ps-short response times.

In the successive steps, we evaluate the functionality of the proposed NL optical response of the studied approaches as neural AF, by emulating their behavior in a 3-layer fully connected NN implemented in the Google Tensorflow tool and for the MNIST dataset. The software implementation MNIST is a known machine-learning dataset comprised of 60,000 grayscale images of handwritten digits. The task is to identify the representing numbers for each image.

In the MNIST dataset each image has 28x28 pixel resolution. In order to feed each image to a fully-connected layer we flatten each 2D image into a vector of 784 pixels.

The first layer of the network is comprised of 100 neurons. As the name suggests, fully-connected networks have all-to-all connections among the neurons in each layer and the incoming inputs. Thus, the first layer requires 784x100 connections. These are identity connections and do not perform any type of weighting to the inputs. The second layer in our network also has 100 neurons, which receive inputs from the first layer. With all-to-all connections, that translates into 100x100 connections. It should be noted that the NL AFs are placed between two consecutive layers on each input connection. As a result, we will have 100x100 NL AFs operating between the first and the second layer. The third layer in our network has only 10 neurons, corresponding to 10 classes of output representing 10 digits from 0 to 9. This last layer can be regarded as a summarizing layer that reduces the dimension of the network output to the 10 required dimensions. For all the layers explained so far, an optical realization is envisioned. However, almost all modern NNs take advantage of a "Softmax" function implemented at the final layer. Softmax converts the incoming input into a probability distribution function with one incoming value receiving a high probability and the remainders receiving much lower probabilities. We have yet to envision an optical realization for this layer, but it is worth noting that this layer can be replaced with similar AFs at the cost of lower accuracy.

We train our network with two sets of AFs, namely, optical AF and commonly used AF in DeepLearning applications. The optical AF are those calculated for MNPs/QD assemblies on top of the waveguide (QD-Top), assemblies in the middle of the waveguide (QD-Middle), and $C_{60}$ films. Respectively, the commonly used software-based AF we used to compare our photonic ones against the Rectified Linear Unit (ReLU), Sigmoid, and Tanh (Fig. 5b) [27]. It is worth mentioning that the actual input and output units of our optical AFs are μWatt. The shown optical AF are for devices of size of 80nm x 30nm. In order to obtain the mathematical model transfer function of the AFs to be used in Tensorflow, we fit quadratic curves to data points from the device simulations, with maximum Root Mean Squared Error (RMSE) due to fitting of 0.23 μWatt. The quadratic AFs are then reconstructed in Tensorflow. It should be noted that the NL AFs acts on the optical power of the incoming electromagnetic radiation associated with the photon flux, which quadratically dependents on the electric field. Moreover, in this study we aim to validate the NN performance by focusing on the shape of the AFs, without taking into account the noise on both system and device level, which will be the subject of an in-depth future investigation. Amplitude and phase noise induced by the NL device, as well as noise in the input signal and weights, could indeed seriously compromise the ability of the AF to discriminate amongst the data, hindering the accuracy of the solution.

We randomly split the MNIST dataset into two subsets of 50,000 and 10,000 images for training and validation respectively. The network was initially set to train for 50 epochs; however, we can see that for QD-middle and $C_{60}$, the model started to over-fit and the training and the validation accuracy dropped. Fig. 5c depicts the accuracy as the network is being trained for 50 epochs, showing that all of the optical AFs are comparable to those of the software-based networks in terms of the accuracy. QD-Middle is the only one performing relatively poorly, converging to a maximum accuracy of 96%. The validation data accuracy corroborates that of training data, as depicted in Fig. 5d. For all the optical NL activation functions a ~100% accuracy is reached, apart from the QD-middle configuration, which reaches a plateau at 99%. It is worth mentioning that the accuracy curves for the validation data are smoother due to the fact that at each validation epoch, the network is evaluated over the whole set of validation images. In contrast, during training we used batch processing and a variation of gradient descent termed Stochastic Gradient Descent (SGD) for training. As a result, at each epoch during the training the network only receives a subset of training images. This results in slight variation of accuracy for different batches of data for different epochs.

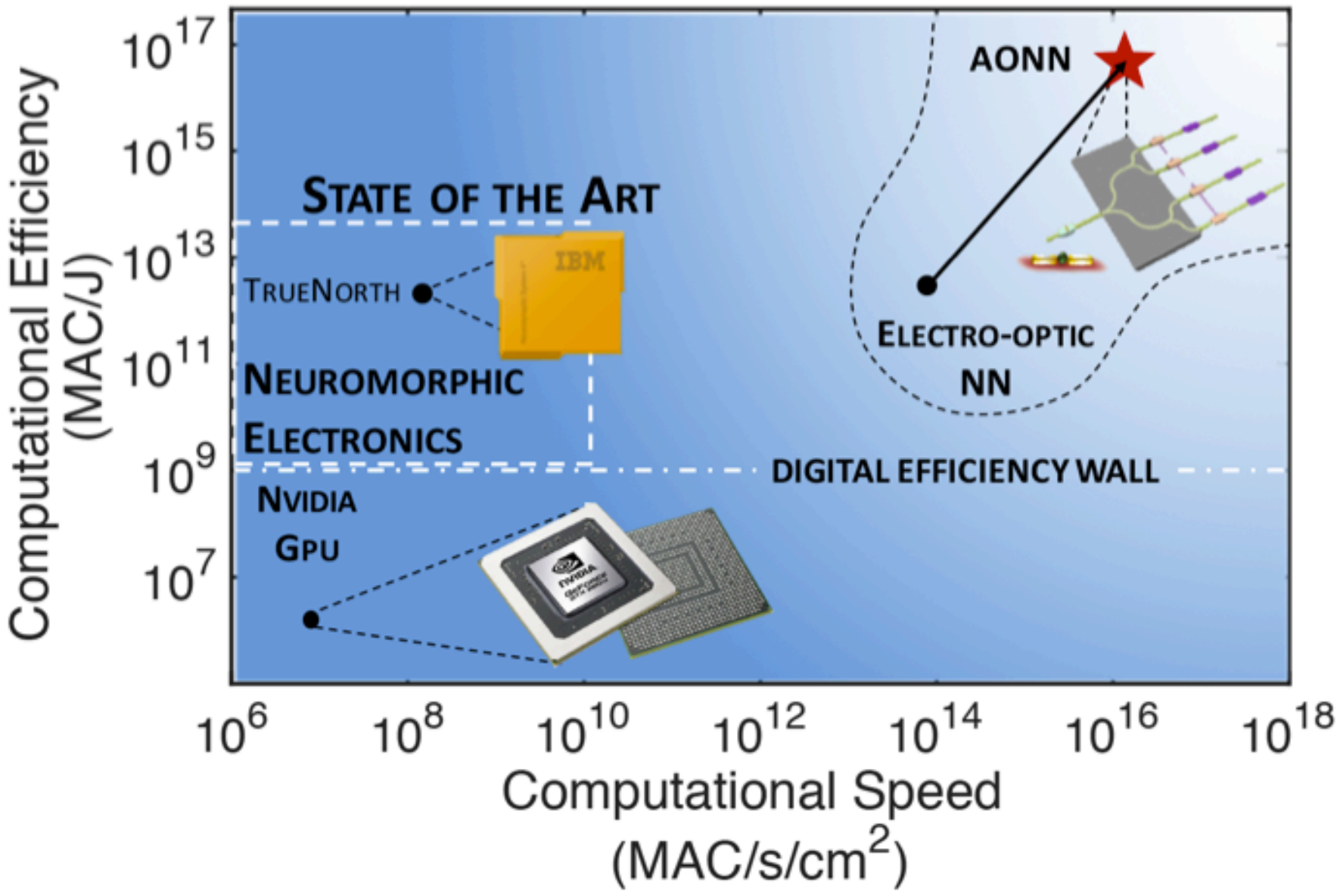

Fig. 6. Comparison of computational energy efficiency and processing speed between existing electronic neuromorphic demonstrations and our proposed programmable photonic platform. NN = neural network, AONN = all-optical NN, GPU = graphical processing unit.

The proposed activation mechanisms based on NL LMI, showed compatible performance, in terms of accuracy as function of epoch with respect to the well-established NL activation functions known from the software-based machine-learning community. The proposed architecture offers possible benefits of the absence of parasitic switching and short delays, since the run-time would simply be given by time-of-flight of a photon through the network. Therefore, an estimate on the processing time can be given by considering the physical length of the components of this photonic integrated circuit NN and its effective waveguide index; for instance the weighted addition obtained through combiners and phase shifters has a physical length of about 100 µm, a passive waveguide synaptic summation has an on-chip coverage of around 200 µm, and an estimated NL activation module is less than 10 µm long. Thus, photonics allows a single neuron to be integrated well within 100's of µm in length, leading to ~ps computation time-scales [28]. This AONN thence would have a delay of about few ps, or $10^{12}$ MAC/s, and $<10^{17}$ MAC/J efficiency given, for example, the power levels of the Fano resonance discussed above. Such performance of the proposed AONN would potentially be several orders of magnitude more efficient and faster than GPU and electro-optical neural networks, provided noise is negligible (Fig. 6) [29].

| ***Technology*** | ***Efficiency (MAC/J)*** | ***Speed (MAC/s)*** |
|---|---|---|
| ***NVIDIA GPU*** *[30]* | *$3x10^6$* | *$10^7$* |
| ***Electro-optical NN*** *[30]* | *$4x10^{12}$* | *$10^{10}$* |
| ***All-optical NN [this work]*** | *$4x10^{16}$* | *$10^{12}$* |

**Table 1.** Comparison of different neuromorphic technologies in terms of computing speed and power efficiency. All optical NN performances are estimated and needs to be considered as an upper-bound.

## 3. Conclusion

In summary, we have investigated a MNPs/QD system, based on two metal NPs sandwiching a QD, which showed a coherent nonlinear optical response. This phenomenon was due to interference between the dipoles of the plasmon oscillation in the metal nanoparticles and the exciton transition in the QD. Furthermore, we modeled integration of the assembly in a waveguide platform and optimized the modulation range of the NLs associated with the transmitted signal, reaching a fully NL optical modulation of the transmitted signal up to 3 dB. Moreover, we also studied the reverse saturable absorption mechanism in a film of $C_{60}$. The film displayed a clear NL optical response as function of the impinging power density, with a modulation range of approximately 7 dB. Moreover, the studied platforms can provide insights into the speed of a complete NN architecture based on integrated photonics and all optical activation functions. The proposed NL optical responses were used as activation functions for fully-connected neural networks, emulated in Tensor Flow. We tested these nonlinear activation functions on a standardized NN training set, MNIST classifiers of handwritten digits. Our results show that the accuracy of the ONN can match others commonly employed for up to 50 number of reconfigurations in both training and validation phase. From an architecture point of view, our all-optical NN has the potential to significantly outperform in terms of computing speed and energy efficiency the established architectures based on either electronics or electro-optics. We estimated an efficiency of $<10^{17}$ MAC/J and a speed of $10^{12}$ MAC/s for a fully optical NN, which is several order of magnitude higher than the electro-optic and FPGA counterparts. These new insights could contribute to the design and fabrication of optical NL modulators, which could pave the way for all-optical high-speed and efficient NNs. Future experimental work is needed to validate this potential.

## 4. Methods

*Simulations*

We use a commercial solver (FDTD Solutions from Lumerical Inc.) built on the finite-difference time-domain method, which solves the Maxwell equations on a discrete spatio-temporal grid, for all the simulations related to the MNPs/QD system to waveguide coupling. The MNPs/QD system is modeled as a 3-D box with the absorptance being swept for modeling the effect of the induced transparency as function of the input power. The adaptive mesh algorithm is used to refine the grid in the QD domain. The RSA model of $C_{60}$ response is realized with a proprietary FDTD-ADE multiphysics code that brings in the carrier kinetics.

## 5. List of Abbreviations

AF (nonlinear) Activation Function of the perceptron

ADE Auxiliary Differential Equation(s)

AONN All-optical Neural Network

CMOS Complementary Metal Oxide Semiconductor

GPU Graphic Process Unit

FDTD Finite Difference Time Domain

FPGA Field Programmable Gate Array

MAC Multiply Accumulate

MNP Metal Nanoparticle

PIC Photonic Integrated Circuit

NP Nano-particle

QD Quantum Dot

NL Nonlinear

NN Neural Network

LMI Light matter Interaction

PNN Photonic Neural Network

RSA Reverse Saturable Absorption

## 6. Funding, acknowledgments, and disclosures

### *6.1 Acknowledgments*

The authors acknowledge fruitful discussion with the team of Prof. Prucnal. S. I. A. and A. V. K. acknowledge the financial support by DARPA/DSO Extreme Optics and Imaging (EXTREME) Program, Award HR00111720032

## 8. Article thumbnail upload

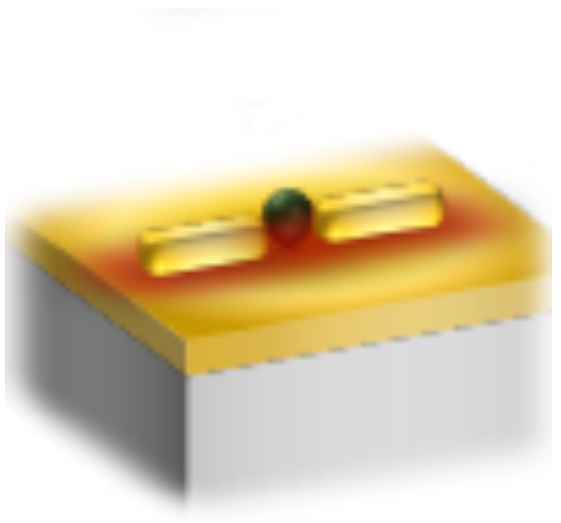

Preview of thumbnail image display on the author submission page.